\documentclass[a4paper, 11pt, twoside]{article}
\usepackage[margin=0.8in]{geometry}

\usepackage{amsmath,amssymb}
\usepackage[english]{babel}
\usepackage{blindtext}
\usepackage{microtype}
\usepackage{graphicx}

\newcommand{\mb}{\boldsymbol}
\newcommand{\mr}{\mathrm}
\newcommand{\pderi}[2]{ \frac{\partial #1}{\partial #2}}

\newcommand{\dotp}[2]{{#1} \cdot {#2}}
\newcommand{\crossp}[2]{{#1}\times{#2}}
\newcommand{\Div}[1]{\nabla \cdot {#1}}
\newcommand{\Curl}[1]{\nabla \times {#1}}
\newcommand{\Grad}[1]{\nabla {#1}}
\newcommand{\Lapla}[1]{\nabla^{2} {#1}}

\newcommand{\JHD}{\mb{J}_{\mr{HD}}}

\begin{document}
\title{Numerical solution of nonlocal hydrodynamic Drude model\\ for
  arbitrary shaped nano-plasmonic structures\\ using N\'ed\'elec finite
  elements}
\author{
Kirankumar~R.~Hiremath$^*$, Lin~Zschiedrich$^+$, Frank~Schmidt$^*$\\
$^*$Computational Nanooptics Group,\\ Konrad-Zuse-Zentrum f\"ur
Informationstechnik Berlin,\\ Takustrasse 7, 14195 Berlin, Germany\\
$^+$JCMwave GmbH, Bolivarallee 22, 14050 Berlin, Germany
}
\date{8 June, 2012}
\maketitle

\begin{abstract}
Nonlocal material response distinctively changes the optical properties
of nano-plasmonic scatterers and waveguides. It is described by the
nonlocal hydrodynamic Drude model, which -- in frequency domain -- is
given by a coupled system of equations for the electric field and an 
additional polarization current of the electron gas modeled analogous
to a hydrodynamic flow. Recent attempt to simulate such nonlocal model
  using the finite difference time domain method encountered difficulties in
dealing with the grad-div operator appearing in the governing equation
of the hydrodynamic current. Therefore, in these studies the model has
been simplified with the curl-free hydrodynamic current approximation;
but this causes spurious resonances. In this paper we present a rigorous weak
formulation in the Sobolev spaces $H(\mathrm{curl})$ for the electric
field and $H(\mathrm{div})$ for the hydrodynamic current, which
directly leads to a consistent discretization based on N\'ed\'elec's
finite element spaces. Comparisons with the Mie theory results
agree well. We also demonstrate the capability of the method to handle
any arbitrary shaped scatterer.
\end{abstract}

\section{Introduction}
Dispersive material properties play an important role in the light-matter
interactions in plasmonic structures. For this quite often the Drude model and
the Lorentz material model are used~\cite{bohren_huffman}, which take
into account spatially purely local interactions between electrons and
the light. In recent investigation it has been found that these local models
are inadequate as the size of the plasmonic scatterers become much
smaller than the wavelength of the incident light~\cite{ruppin_01,
  palomba_08}. To overcome this, a sophisticated nonlocal material model 
is required, such as the hydrodynamic model of the electron gas as discussed by
Boardman \textit{et~al.}~\cite{boardman_82}.

In the first principle formulation, the hydrodynamic model of the electron gas
is formulated by coupling macroscopic time domain Maxwell's equations for
electromagnetic fields with the equations of motion of the
electron gas which behaves similar to hydrodynamic
flow~\cite{boardman_82}. This gives rise to a hydrodynamic polarization
current.  The resultant 
coupled system of equations is flexible enough to incorporate a variety of
advanced quantum mechanical effects. When considered only the kinetic
energy of the free electrons, it yields the nonlocal hydrodynamic Drude
model (discussed in Sec.\,\ref{sec:HDM}).

In one of the earlier attempts, the nonlocal hydrodynamic Drude model has been simulated
with the finite difference time domain (FDTD) method, but with the
quasi-static approximation~\cite{mcmahon_10}.  As a consequence of this
approximation, the tensorial grad-div operator ($ 
\nabla (\nabla \cdot \mb{A})$) appearing in the governing equation for
the hydrodynamic current simplifies to vectorial linear Laplacian
operator ($ \nabla^{2} \mb{A}$). This was needed to render the
system into a form suitable for the standard FDTD framework. However the
comparison with the analytical Mie theory~\cite{ruppin_01} showed that
this approach produces spurious plasmonic resonances below the plasma
frequency~\cite{raza_11}.

\begin{figure}[!htb]
  \centering
\includegraphics[width=0.3\textwidth]{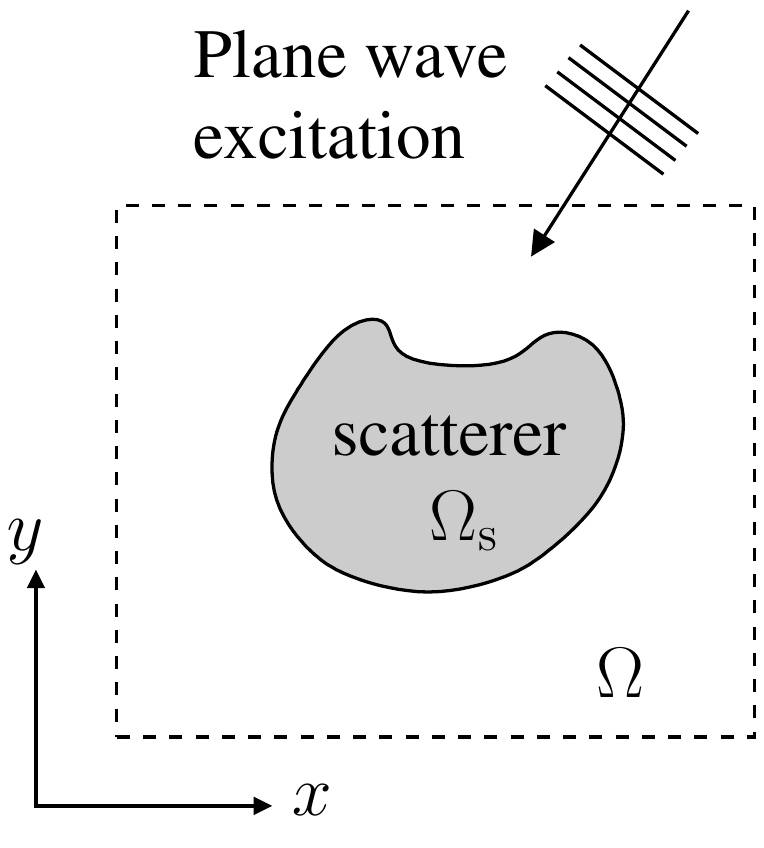}
  \caption{Scattering setting: A plane wave of angular frequency $\omega$
is incident on a nano-plasmonic scatterer with nonlocal material
properties and arbitrary shaped domain $\Omega_{\mr{s}}$. The scattering problem is defined on the entire space, but for numerical computations
we restrict it to a finite (computational) domain $\Omega$, which
necessitates construction of transparent boundary conditions
(e.g. perfectly matched layer). We assume a homogeneous medium outside
$\Omega$.}  
  \label{fig:setting}
\end{figure}

In this paper we do not rely on the quasi-static approximation, and we
present a rigorous weak formulation in the frequency domain (with time
dependence $\exp{(-i \omega t)}$ and for a typical light scattering
setting as shown in Fig.\,\ref{fig:setting}), which directly allows a consistent
discretization within N\'ed\'elec finite element
spaces. We would like to point out that while the
  present work was under review, Toscano \textit{et al.} have independently
  reported a finite element approach for the simulation of the nonlocal
  hydrodynamic Drude model~\cite{toscano_12}. While the emphasis of
  their work is on analyzing physical effects due to the nonlocality,
  the main contribution of this work is to present appropriate finite
  element framework behind such computational scheme (see
  Sec.\,\ref{sec:wp}). This will ensure that the finite element
  solutions are physically meaningful.

The paper is organized as follows. Sec.\,\ref{sec:HDM} introduces the
nonlocal hydrodynamic Drude model, various approximations involved
in it, and derives the governing coupled system of equations for the
electric field and the nonlocal hydrodynamic current. A finite element
formulation for the governing equation of the electric
field follows the standard procedure based on the curl conforming elements
(as in the case of the usual local material models~\cite{bondeson_05}), but
one needs to choose an appropriate finite element space for solution
of the equation for the hydrodynamic current. We discuss this consistent weak
formulation  and the finite element setting of the model in
Sec.\,\ref{sec:wp}. Then in Sec.\,\ref{sec:ne} we validate the 
simulation results of this model with analytical Mie theory
results. We also demonstrate the ability of the method to handle arbitrary shaped geometry with the example of V groove channel
plasmon-polariton devices.

\section{Nonlocal hydrodynamic Drude model}
\label{sec:HDM}
For the sake of clarity and to highlight various approximations involved
in the nonlocal hydrodynamic Drude model, we derive its governing
equations in this section. The macroscopic Maxwell's equations for
non-magnetic materials with no external current density density and no
external charge density are 
  \begin{eqnarray}
 \nabla \times \mb{E}(\mb{r},\omega)  & =   & i \omega \mu_{0}
 {\mb{H}}(\mb{r},\omega),  \label{eq:FDcurlE} \\
     \nabla \times \mb{H}(\mb{r},\omega)  & = & -i \omega
     \varepsilon_{0} \varepsilon_{\mr{loc}}(\mb{r}, \omega) 
     \mb{E}(\mb{r},\omega) + \JHD(\mb{r}, \omega), \label{eq:FDcurlH}\\ 
      \nabla \cdot \mb{D}(\mb{r},\omega) & = & -e n(\mb{r},
      \omega), \label{eq:FDdivD} \\ 
      \nabla \cdot \mb{B}(\mb{r},\omega) & = & 0,
\end{eqnarray}
where $\mb{E}$ is the electric field, $\mb{H}$ is the magnetic field,
$\mb{D} = \varepsilon_{0} \varepsilon_{\mr{loc}} \mb{E}$ is the electric displacement, and $\mb{B} = \mu_0 \mb{H}$ is
the magnetic induction.   $\varepsilon_{0}$ is the permittivity
constant and $\mu_{0}$ is the permeability constant. A part of the relative permittivity due to the
local-response is defined as $\varepsilon_{\mr{loc}} = \varepsilon_{\infty}(\mb{r})  +
\varepsilon_{\mr{inter}}(\mb{r}, \omega)$. $\varepsilon_{\infty}$ is the relative permittivity for infinite
frequency. If the material system under consideration has interband
transitions, then the corresponding relative permittivity (typically
described by the Lorentz material model) is given by
$\varepsilon_{\mr{inter}}$.

The effect of nonlocal material response on plasmonic scatterers is
incorporated by an auxiliary nonlocal hydrodynamic
current density $\JHD$. This polarization current is defined only 
on the spatial domain $\Omega_{\mr{s}}$ where the plasmonic scatterer exists, and set to
zero outside it. The exact material interface conditions for  $\JHD$ are discussed later on. The current density $\JHD$
is related with the electron gas density $n$ by
 \begin{equation}
    \label{eq:J1}
    \JHD(\mb{r}, \omega) = -e n(\mb{r}, \omega) \mb{v}(\mb{r}, \omega),
  \end{equation}
where $e$ is the charge of the electron. In time domain, the
hydrodynamic velocity $\mb{v}$ is related with the electron gas density
$n$ by the continuity equation $ \pderi{n(\mb{r}, t)}{t} +
\Div{(n(\mb{r}, t)\mb{v}(\mb{r}, t) )}= 0$. Let $n_{0}$ be the constant
equilibrium density of the electron gas, and $n_{1}$ be the linear  
perturbation, then the time dependent density $n$ is
written as $n(\mb{r}, t)  = n_{0} + n_{1}(\mb{r}, t)$. Note that $n_{0}$
and $n_{1}$ are nonzero only in $\Omega_{\mr{s}}$. With this
linearization ansatz, a term $\Div{(n_{1}(\mb{r}, t)\mb{v}(\mb{r}, t))}$ is
negligible, and the continuity equation in frequency domain simplifies to
  \begin{equation}
    \label{eq:ceq}
    -i \omega n_{1}(\mb{r},
    \omega)  + n_{0} \Div{\mb{v}(\mb{r},\omega)}  = 0.
  \end{equation}
The hydrodynamic velocity $\mb{v}(\mb{r}, \omega)$ obeys the generalized
momentum equation derivable from quantum mechanical
Hamiltonian~\cite{boardman_82}, which in frequency domain is given by
\begin{equation}
  \label{eq:Ham}
  m_{\mr{e}} \left ( -i \omega + \dotp{\mb{v}}{\nabla} \right ) \mb{v} =
  -e(\mb{E} + \crossp{\mb{v}}{\mb{B}}) -  m_{\mr{e}} \gamma \mb{v} -
  \nabla \left( \frac{\delta g[n]}{\delta n} \right),
\end{equation}
where $m_{\mr{e}}$ is the effective electron mass, $\gamma$ is the
damping constant (= inverse of the collision time) and $g[n]$ is energy
functional of the fluid.

Several further approximations are introduced in order to deal with
Eq.\,\eqref{eq:Ham}. Assume that the nonlinear
term corresponding to the hydrodynamic total derivative
$(\dotp{\mb{v}}{\nabla}) \mb{v}$ is negligible. Also, the driving force
of the electron fluid is only the electric field $\mb{E}$, and therefore
we neglect the effect of the magnetic induction field $\mb{B}$. For the
free electron gas, assume only the kinetic energy constitute $g[n]$
(neglecting the exchange and the correlation effects)~\cite{boardman_82}, and
along with Eq.\,\eqref{eq:ceq} one can estimate 
\begin{equation}
  \Grad{\left(\frac{\delta g[n]}{\delta n} \right )}  \approx  m_{\mr{e}}
  \beta^{2} \frac{1}{n_{0}} \Grad{n_{1}}
= m_{\mr{e}} \beta^{2} \frac{1}{i \omega} \Grad{(\Div{\mb{v}(\mb{r},
    \omega)})}, \label{eq:g1} 
\end{equation}
with $\beta^{2} = \frac{3}{5} v_{\mr{F}}^{2}$ is a term
proportional to the Fermi velocity $v_{\mr{F}}$ (here the value of the
constant of proportionality is taken as $3/5$, but to be precise, it
depends on the various properties of the physical setting under
consideration~\cite{boardman_82}).

With these approximations Eq.\,\eqref{eq:Ham} gives
\begin{eqnarray*}
\beta^{2} \Grad{(\Div{\mb{v}(\mb{r}, \omega)})} 
+ \omega (\omega + i \gamma) \mb{v}(\mb{r}, \omega) 
& = & -i \omega \frac{e}{m_{\mr{e}}} \mb{E}(\mb{r}, \omega).
\end{eqnarray*}

Multiplying this equation by  $-e n_{0}$, we get the governing equation
for the nonlocal hydrodynamic current density $\JHD$
\begin{eqnarray}
\beta^{2} \Grad{(\Div{\JHD(\mb{r}, \omega)})} 
+ \omega( \omega +  i\gamma) \JHD(\mb{r}, \omega) 
& = &  i \omega \omega_{\mr{p}}^{2} \varepsilon_{0} \mb{E}(\mb{r}, \omega),  
\label{eq:FDHD_J}~
\end{eqnarray}
where $\omega_{\mr{p}}^{2} = \frac{e^{2} n_{0}}{\varepsilon_{0}
  m_{\mr{e}}}$ is the plasma frequency of the free electron gas. In this equation the macroscopic electric field $\mb{E}$ acts as a
source for evolution of the hydrodynamic current. In turn, this hydrodynamic
current influences the evolution of the electric field $\mb{E}$. This
part of the model is obtained by taking curl of \eqref{eq:FDcurlE}, and
using \eqref{eq:FDcurlH}, and then rearranging, we get the familiar curl-curl
equation for $\mb{E}$ as
  \begin{equation}
    \nabla \times \mu_{0}^{-1}( \nabla \times \mb{E}(\mb{r}, \omega)) - \omega^{2}
\varepsilon_{0} \varepsilon_{\mr{loc}}(\mb{r}, \omega) 
\mb{E}(\mb{r}, \omega)  =   i \omega \mb{J}_{\mr{HD}}(\mb{r}, \omega).
\label{eq:FDHD_E}
  \end{equation}

Eq.\,\eqref{eq:FDHD_J} and \eqref{eq:FDHD_E} are the required coupled system
of equations for the nonlocal hydrodynamic Drude
model. Eq.\,\eqref{eq:FDHD_E} is defined on unbounded domain, but for
numerical computations,  it is restricted to a finite computational
domain $\Omega$ by the transparent boundary condition like the perfectly matched
layer. Whereas Eq.\,\eqref{eq:FDHD_J} is solved 
on the region $\Omega_{\mr{s}}$ containing the material with the nonlocal
response, and outside it $\JHD=0$. Since the normal component of   $\JHD$ is
continuous across the material interfaces, it leads to
$\dotp{\mb{n}}{\JHD} = 0$ on the material interfaces.\\

\noindent \textbf{Curl-free approximation:} Using the vector calculus
identity $ \Grad{(\Div{\mb{A}})} = \Curl{\Curl{\mb{A}}} +
\Lapla{\mb{A}}$, and assuming $\Curl{\JHD} = 0$, Eq.\,\eqref{eq:FDHD_J}
becomes
\begin{equation}
 \beta^{2}  \Lapla{\JHD}(\mb{r}, \omega) + \omega( \omega +  i\gamma) \JHD(\mb{r}, \omega) 
 =   i \omega \omega_{\mr{p}}^{2} \varepsilon_{0} \mb{E}(\mb{r},
 \omega). \nonumber
\end{equation}
This is precisely  the frequency domain representation of the corresponding
time domain equation, which is solved in Ref.\,\cite[Eq.\,14]{mcmahon_10}.


\section{Weak formulation}
\label{sec:wp}

In this section we bring the light scattering problem of
Fig.\,\ref{fig:setting} into a variational form. We start with
Eq.\,\eqref{eq:FDHD_E} for the electric field, for which
  the weak formulation follows the standard procedure based on
  N\'ed\'elec's curl conforming finite elements~\cite{bondeson_05,
    nedelec_86}. An appropriate ansatz  space for the electric field is
the Sobolev space  
\begin{equation}
H(\mr{curl}, \Omega)=\left \{\mb{E} \in (L^2(\Omega))^3\,|\, \Curl{\mb{E}} \in
(L^2(\Omega))^3 \right \}, \nonumber
\end{equation}
which contains fields with weakly defined curl-operator defined on the
domain $\Omega$~\cite[Sec. 3.5]{monk_03}. 

Multiply Eq.\,\eqref{eq:FDHD_E} with a trial function $\varphi \in H(\mr{curl}, \Omega)$, and integrate over $\Omega$.  Then partial integration yields
\begin{equation}
\label{eq:FDHD_WE0}
  \int_\Omega  \left( (\Curl{\varphi}) \cdot (\mu_{0}^{-1}
    \Curl{\mb{E}}) - \omega^{2} 
\varphi \cdot \varepsilon_{\mr{loc}}
\mb{E}\right)\,dV   
+\int_{\partial \Omega} \varphi \cdot (\mb{n} \times  (\mu_{0}^{-1} 
\Curl{\mb{E}}))\,dA
 =  i \omega \int_\Omega \varphi \cdot \mb{J}_{\mr{HD}}\,dV,
\end{equation}
with the local permittivity $\varepsilon_{\mr{loc}}$ and with the outer
normal $\mb{n}$ of the computational domain. At this stage, we encounter
the problem of defining boundary conditions of the electric field on
$\partial \Omega$, which is addressed by the transparent boundary
condition. This can be realized in various forms like perfectly matched
layers, infinite element method, etc.~\cite[Ch.\,13]{monk_03}; but here,
for the notational simplicity we will make use of the Dirichlet to Neumann (DtN)
operator~\cite{zschiedrich_09} (also known as the Calderon map
approach~\cite[Sec.\,9.4]{monk_03}).

Outside the scatterer the electric
field is a superposition of the exciting (incoming) field
$\mb{E}_{\mr{inc}}$ and the {\em scattered} field $\mb{E}_{\mr{s}},$
i.e. $\mb{E}=\mb{E}_{\mr{inc}}+\mb{E}_{\mr{s}}.$ The outward radiating
scattered field satisfies Maxwell's equations in the exterior domain, and
the Silver-M\"uller radiation condition at infinity. But then $\mb{E}_{\mr{s}}$ is already defined by it's Dirichlet field values on $\partial \Omega.$ Especially, one is able to determine the Neumann field values $ \mb{n} \times (\mu_{0}^{-1} \Curl{ \mb{E}_{\mr{s}}})|_{\partial \Omega}$ from $\mb{E}_{\mr{s}}|_{\partial \Omega}$. This mapping defines the so called DtN-operator. For the above Neumann boundary term we get
\begin{eqnarray*}
\int_{\partial \Omega} \varphi \cdot (\mb{n} \times  (\mu_{0}^{-1} 
\Curl{\mb{E}}))\,dA & = & \int_{\partial \Omega} \varphi \cdot (\mb{n} \times (\mu_{0}^{-1}\Curl{(\mb{E}_{\mr{inc}}+\mb{E}_{\mr{s}}) }))\,dA \\
{ } & = &  \int_{\partial \Omega} \varphi \cdot (\mb{n} \times  (\mu_{0}^{-1}\Curl{\mb{E}_{\mr{inc}} }))\,dA + \int_{\partial \Omega} \varphi \cdot  \mr{DtN} ( \mb{E}_{\mr{s}})\,dA.  
\end{eqnarray*} 
Using $\mb{E}=\mb{E}_{\mr{inc}}+\mb{E}_{\mr{s}}$ once more, we can eliminate the scattered field and recast Eq.\,\eqref{eq:FDHD_WE0} to
\begin{eqnarray}
\label{eq:FDHD_WE1}
  \int_\Omega  ((\Curl{\varphi}) \cdot (\mu_{0}^{-1} \Curl{\mb{E}}) - \omega^{2}
\varphi \cdot \varepsilon_{\mr{loc}}
\mb{E})\,dV  
& + & \int_{\partial \Omega} \varphi \cdot  \mr{DtN} ( \mb{E}) \,dA -
i \omega \int_\Omega \varphi \cdot \mb{J}_{\mr{HD}}\,dV \nonumber
  \\
= -  \int_{\partial \Omega} \varphi \cdot (\mb{n} \times
(\mu_{0}^{-1}\Curl{\mb{E}_{\mr{inc}} }))\,dA & + & 
\int_{\partial \Omega} \varphi \cdot  \mr{DtN} (
\mb{E}_{\mr{inc}})\,dA,  \;\forall~\varphi \in
H(\mr{curl}, \Omega),  {}  {}
\end{eqnarray}
where only the exciting field $\mb{E}_{\mr{inc}}$ appears on the right hand side. 
 
It remains to bring the Eq.\,\eqref{eq:FDHD_J} for the hydrodynamic
current into variational form. The subsequent weak
  formulation reveals that for a  
physically meaningful solution, the required ansatz space for $\JHD$
needs to be divergence conforming, which can be precisely realized by the
N\'ed\'elec's divergence conforming finite
elements~\cite{nedelec_86}. Thus the appropriate ansatz space for weak
formulation of Eq.\,\eqref{eq:FDHD_J} is the Sobolev space 
\begin{equation}
H_0(\mr{div}, \Omega_\mr{s})=\left \{\JHD  \in
(L^2(\Omega_{\mr{s}}))^3\,|\, \Div{\JHD} \in
(L^2(\Omega_{\mr{s}}))^3,\;\mb{n} \cdot \JHD =0\; \mbox{on}\;\partial
\Omega_\mr{s} \right \}. \nonumber
\end{equation}
This restricts the hydrodynamic current to the plasmonic scatterer, and
imposes zero normal component on the boundary of the scatterer. This
reflects the physical requirement that the nonlocal hydrodynamic electron gas is not allowed to flow out of the scatterer. Then the variational form of Eq.\,\eqref{eq:FDHD_J} reads as
\begin{equation}
- \int_{\Omega_{\mr{s}}} \beta^{2} (\Div{\psi}) (\Div{\JHD})\, dV 
+ \omega( \omega +  i\gamma) \int_{\Omega_{\mr{s}}}  \psi \cdot \JHD \, dV
-i \omega \omega_{\mr{p}}^{2}  \int_{\Omega_{\mr{s}}} \psi \cdot
\epsilon_{0} \mb{E} \, dV =  0,~\hfill
\forall~\psi  \in   H_{0}(\mr{div}, \Omega_{\mr{s}}). \label{eq:FDHD_WJ}
\end{equation}

After the problem is formulated  in the Sobolev space $H(\mr{curl},
\Omega) \times H_0(\mr{div}, \Omega_\mr{s})$ for $(\mb{E}, \JHD)$,
one can use N\'ed\'elec finite element spaces, which lead to a consistent
discretization of the problem, fulfilling the required boundary and
material interface conditions~\cite[Ch. 5]{monk_03}.

\section{Numerical examples}
\label{sec:ne}
Although the above weak formulation and the N\'ed\'elec elements based
finite element method is discussed for a full 3D setting, for the sake
of simplicity we restrict ourselves to a 2D setting (in the $XY$ plane)
for numerical illustrations. Here the incident plane wave is either
s-polarized (i.e. out-of-plane in the $z$-direction) or p-polarized
(i.e. in the $XY$ plane). Since for the above 2D
  settings the s-polarized source can not excite plasmonic effects, we
consider only the p-polarized incident field.

Accuracy and efficiency of the numerical solution of
  the nonlocal hydrodynamic model depends on implementation of the
  transparent boundary condition in Eq.\,\eqref{eq:FDHD_WE0}. Here we
  benefit from the in-house developed finite element code
  JCMsuite~\cite{zschiedrich_09, pomplun_07}. We have observed that
  solving the resultant discrete coupled system of equations iteratively
  as in Ref.\,\cite{toscano_10} causes slow convergence and numerical
  issues; therefore we solve it directly with a sparse LU
  decomposition.

\subsection{Cylindrical plasmonic nanowires}
\label{sec:cnw}
For a validation of the present approach, we simulate cylindrical nanowires. Extending the Mie theory for the nonlocal response,
Ruppin had formulated the analytical solution for this
problem~\cite{ruppin_01}. When this setting was simulated with the
curl-free hydrodynamic current approximation as in
Ref.\,\cite{mcmahon_10}, spurious (model induced) resonances were
produced, which has been discussed in detail in
Ref.\,\cite{raza_11}. Thus the cylindrical nanowire serves as a good
benchmark problem.

As in Ref.\,\cite{ruppin_01}, the cylindrical nanowire is of radius
$R=2$ nm, and is made up of a dispersive material with
$\epsilon_{\infty} = 1$ (and no interband transitions), plasma
frequency $\omega_{\mr{p}}=8.65 \times 10^{15}$ s$^{-1}$, damping
constant $\gamma = 0.01 \omega_{\mr{p}}$. The system constant $\beta^{2} =
\frac{3}{5} v_{\mr{F}}^{2}$ is computed for the Fermi velocity
$v_{\mr{F}}=1.07 \times 10^{6}$ ms$^{-1}$. The nanowire placed in the
exterior medium of refractive index 1, and is excited
with a unit amplitude, $y$-polarized plane wave propagating in the direction
of $x$-axis. With these parameters the coupled system of equations
\eqref{eq:FDHD_WE1} and \eqref{eq:FDHD_WJ} are solved. 

\begin{figure}[!htb]
  \centering
\includegraphics[width=0.5\textwidth]{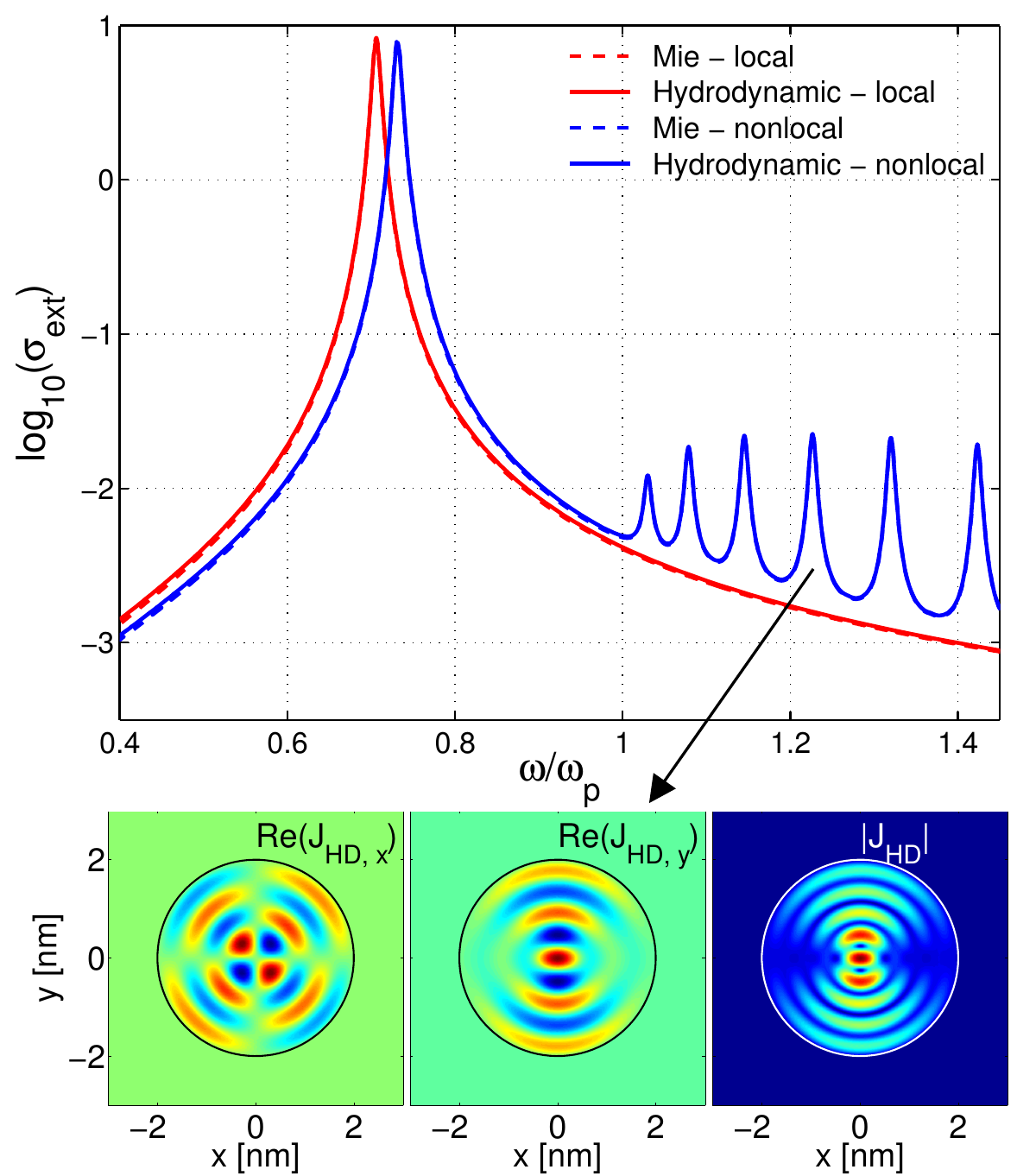}
  \caption{Simulation results for the cylindrical nanowire in Sec.\,\ref{sec:cnw}.  The curves show
    comparison of the finite element numerical solutions for the 
    nonlocal (solid blue line) and the local (solid red line) hydrodynamic
    model with the corresponding analytical solutions (dashed blue line
    and dashed red line respectively) based on the Mie theory. The field
    plots show simulated hydrodynamic current $\JHD$ corresponding to
    the forth order nonlocal resonance at $\omega/\omega_{\mr{p}} =
    1.227$ (indicated by the arrow).} \label{fig:comp_HD_Mie}
\end{figure}

In this frame-work, we can simulate the conventional local Drude model
by explicitly breaking the hydrodynamic coupling by setting $\JHD \equiv
0$, and using the  local Drude material model for the local relative
permittivity $\varepsilon_{\mr{loc}}$ in Eq.\,\eqref{eq:FDHD_WE1}. Following the
conventions in Ref.\,\cite{ruppin_01}, we compute the normalized
extinction cross section $\sigma_{\mr{ext}}$ (the
usual extinction cross section normalized by the
diameter of the cylindrical wire). Fig.\,\ref{fig:comp_HD_Mie} shows
the results plotted for 
the normalized angular frequency $\omega/\omega_{\mr{p}}$ (normalized with
respect to the Drude plasma frequency $\omega_{\mr{p}}$).

The finite element numerical solution for the nonlocal hydrodynamic model
(solid blue line) has a prominent peak at $\omega/\omega_{\mr{p}} =
0.731255$, which corresponds to the localized surface plasmon resonance. The subsidiary
peaks beyond the bulk plasma frequency (e.g. at
$\omega/\omega_{\mr{p}} = 1.03002$, $1.07888$, $1.14547$, $1.22707$, $\cdots$
etc.) are due to the nonlocal hydrodynamic current. Consistent with the
observations in Ref.\,\cite{ruppin_01}, these peaks are present beyond the bulk
plasma frequency. The positions of the  localized surface plasmon resonance and the
nonlocal hydrodynamic Drude resonances agree well with the
analytical Mie results (the blue dashed line, which is largely covered by the 
solid blue line). Good agreement has been also observed in case of
the local Drude model simulations; where the main peak due to the
surface plasmon resonance has been shifted towards lower frequency
($\omega/\omega_{\mr{p}} = 0.706086$, which is quite close to the
theoretical estimation of $\omega/\omega_{\mr{p}} = 1/\sqrt{2}  
= 0.70711$).

The subplots in Fig.\,\ref{fig:comp_HD_Mie} show the simulated current
density $\JHD$  for the nonlocal hydrodynamic resonance at
$\omega/\omega_{\mr{p}} =1.227$. Non curl-free nature of these field
plots clearly show that the quasi-static approximation (as in
Ref.\,\cite{mcmahon_10}) is inaccurate.

\subsection{Plasmonic V groove channel plasmon-polariton
  devices }
\label{sec:vg}
Having verified the method for the test case of cylindrical nanowires, now
we demonstrate the capability of the method to handle an arbitrary
shaped geometry. One of such geometries which is of a great practical
interest is a channel plasmon-polariton (CPP) devices with a V
groove~\cite{bozhevolnyi_06}. Modal and resonance properties of such
devices  have been investigated thoroughly. Accurate numerical
simulation of the V groove geometry is challenging due to subwavelength
device  features and field enhancement due to plasmonic
effects~\cite{moreno_06, hafner_07}. This makes the V groove geometry
an interesting candidate to check the effect of the nonlocal response. 

\begin{figure}[!hbt]
  \centering
\includegraphics[width=0.88\textwidth]{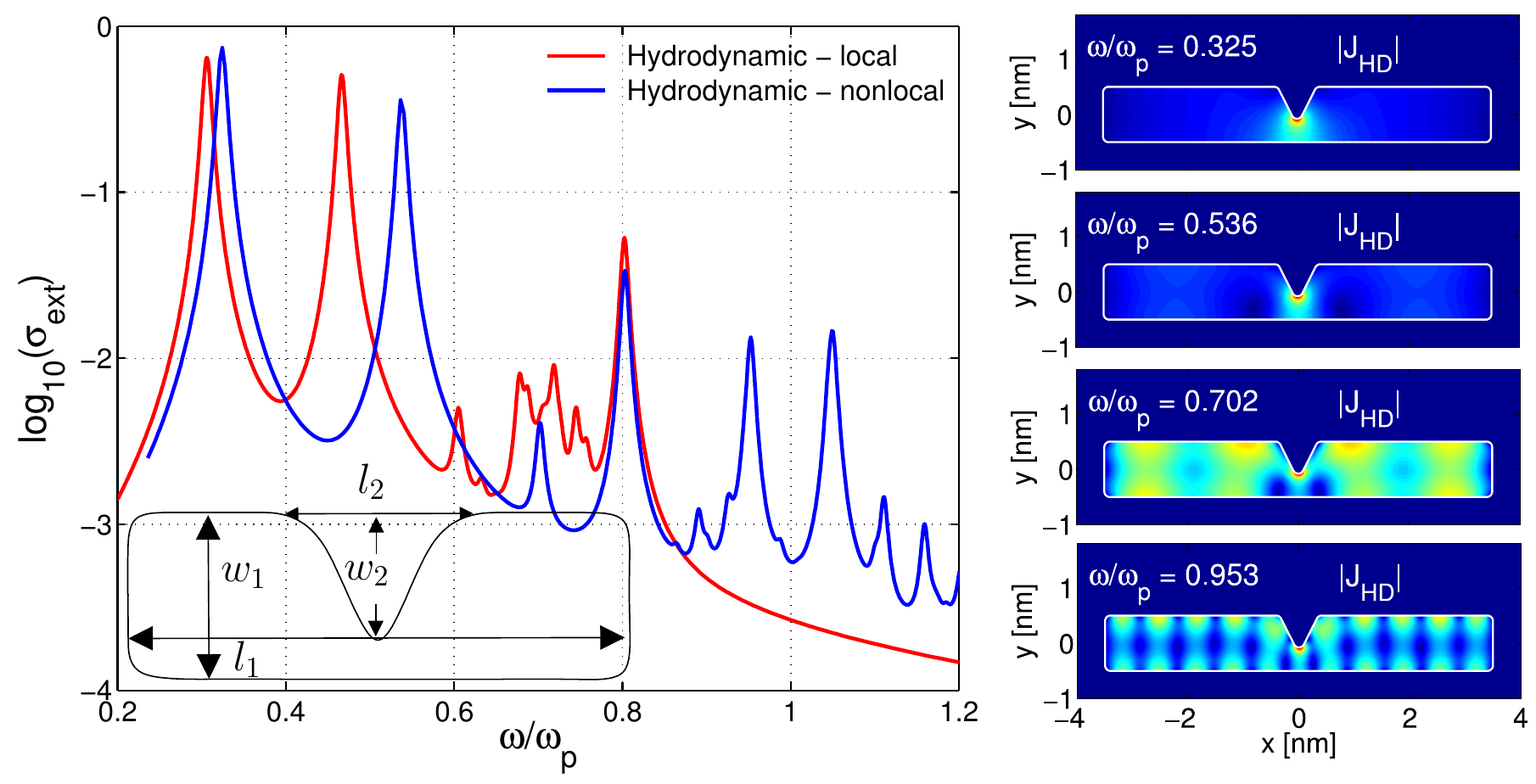}
  \caption{Left: Effect of the nonlocal material response on the resonance
    modes of V groove channel plasmon-polariton structures. The device
    parameters are as in Sec.\,\ref{sec:vg}. Right: Field distribution
    for the hydrodynamic current $\JHD$ for various nonlocal resonances
    of the V groove.}  
  \label{fig:comp_groove}
\end{figure}

We simulate scattering off a V groove configuration of length $l_{1} =
7$ nm, width $w_1 = 1$ nm, with a symmetrically placed groove of length
$l_{2} = 0.7$ nm, width $w_2 =  0.7$ nm. As shown in the inset of
Fig.\,\ref{fig:comp_groove}, the sharp corners of the  device  are
rounded with the corner radius of $0.1$ nm. Note that by decreasing the
corner rounding radius, the response of the V groove device will
slightly change quantitatively, but we have made sure that the
qualitative features are not changed significantly. The material and the
hydrodynamic parameters are taken as in the case of cylindrical
nanowires in Sec.\,\ref{sec:cnw}. Resonance modes of the this device are
excited by a unit amplitude, $x$-polarized (parallel to the length of
the device ) plane wave propagating in the direction of minus $y$-axis
(parallel to the width of the device). As in Sec.\,\ref{sec:cnw}, we 
again analyze the normalized extinction cross section
$\sigma_{\mr{ext}}$ for the V groove structure, but now the
normalization is done with the length of the device  $l_{1}$.
 
First we simulated the above V groove structure
   for the local Drude material model
  (the conventional model). As seen from the red curve in Fig.\,\ref{fig:comp_groove},
several plasmon-polariton resonances are excited. The most interesting features
are the resonances corresponding to the prominent peaks at
$\omega/\omega_{\mr{p}} = 0.306$ and $\omega/\omega_{\mr{p}}
= 0.466$, for which high electric field intensity is present in the
narrow groove region~\cite{hafner_07}. On the other hand, for the
resonance at $\omega/\omega_{\mr{p}}
= 0.8$, the electric field is localized at the outer vertical
edges of the V 
groove structure, and no such particular distinction can be made for the
other less prominent resonance states (e.g. the ``local'' resonances
around $\omega/\omega_{\mr{p}} = 0.7$).

When this V groove structure is simulated for the nonlocal Drude
material model, the resonance spectrum changes significantly (the blue
curve). Similar to the case of the nanowires, the both local
Drude model plasmon-polariton resonances (corresponding to the high
field localization in the groove) experience a shift towards high
frequency ($\omega/\omega_{\mr{p}} = 0.325$ and $\omega/\omega_{\mr{p}}
= 0.536$), but the individual extents of these shifts are 
different. The other local resonances are also influenced by the
hydrodynamic current, resulting in  high order nonlocal hydrodynamic
resonances. To get a closer look at these resonances, field 
distributions of the hydrodynamic current are illustrated in  
Fig.\,\ref{fig:comp_groove}. For the resonances below the the plasma 
frequency, for example at $\omega/\omega_{\mr{p}} = 0.702$ and
$\omega/\omega_{\mr{p}} = 0.953$, these field plots show the oscillating
hydrodynamic current; which indicates that unlike as in the case of the
nanowires, for the V groove structures the plasma frequency
$\omega_{\mr{p}}$ does not seem to separate the high order nonlocal
hydrodynamic resonances from the plasmon-polariton 
resonances. Since the ``local'' resonances around
$\omega/\omega_{\mr{p}} = 0.7$ do not show any distinctive field
localization properties, it is difficult to associate them
with the high order nonlocal resonances.  For the present simulation
setting, some of these nonlocal hydrodynamic
resonances are more prominent than the minor local
resonances. It gives the   indication with the inclusion of nonlocal
effect, the resonance properties   of the CPP devices
change significantly.

\section{Conclusions}
In this work we discussed a weak formulation for the nonlocal
hydrodynamic Drude model, which is simulated with N\'ed\'elec finite
element method. Unlike the previously reported work~\cite{mcmahon_10},
this approach does not use the curl-free approximation, and thus avoids
spurious (i.e. model or approximation induced) resonances. The simulated
results agree well with the analytical results based
on the Mie theory, 
and the method is capable of handling arbitrary shaped scatterers. The
approach discussed in this work will serve as a reference for
investigating advanced hydrodynamic models, which will take into account
additional physical effects.

\section*{Acknowledgments}
This work is partially funded by the DFG (German Research Foundation)
priority program 1391 ``Ultrafast Nanooptics''. The authors are thankful
to Sven Burger (Zuse Institute, Berlin) for useful discussions and help
for simulations.


\end{document}